\documentclass[aps,pra,floatfix,twocolumn,showpacs]{revtex4-1}
\usepackage{graphicx}
\usepackage{amsmath}
\usepackage{amsfonts}
\usepackage{amssymb}
\usepackage{epsfig}
\usepackage[pdftex]{color}
\usepackage{amsmath,graphicx,amssymb,braket,xcolor,subfigure,upgreek}
\usepackage[colorlinks, linkcolor=blue, citecolor=blue, urlcolor=blue, breaklinks=true]{hyperref}
\usepackage{float}

\begin{document}

\title{Light-matter interactions in multi-element resonators}

\author{Claudiu Genes$^{1}$ and Aur\'{e}lien Dantan$^2$}

\affiliation{$^1$ Max Planck Institute for the Science of Light, 91058 Erlangen, Germany \\
$^2$ Department of Physics and Astronomy, University of Aarhus, DK-8000 Aarhus C, Denmark}

\date{\today}

\begin{abstract}
We investigate light-matter interactions in multi-element optical resonators and provide a roadmap for the identification of structural resonances and the description of the interaction of single extended cavity modes with quantum emitters or mechanical resonators. Using a first principle approach based on the transfer matrix formalism we analyze, both numerically and analytically, the static and dynamical properties of three- and four-mirror cavities. We investigate in particular conditions under which the confinement of the field in specific subcavities allows for enhanced light-matter interactions in the context of cavity quantum electrodynamics and cavity optomechanics.
\end{abstract}



\maketitle

\section{Introduction}

Cavity quantum electrodynamics (QED) provides a playground for testing and controlling the interaction of quantized electromagnetic fields with matter at both the micro- (individual quantum emitters) and the macro-scale (suspended or levitated reflectors). The remarkable progress in this field during the past decades has lead to, among others, the observation and manipulation of the coherent dynamics of individual quantum systems~\cite{Haroche2006}, the realization of efficient quantum light-matter interfaces~\cite{Kimble2008}, the exploration of many-body dynamics of cold atoms in cavity-generated potentials~\cite{Ritsch2013} or improved cooling and control of mechanical nanoresonators~\cite{Aspelmeyer2014}.

A paradigm for cavity QED is a system of two highly-reflecting mirrors in a linear Fabry-Perot arrangement. The confinement of the quantized electromagnetic modes between the mirrors allows for the independent treatment of single cavity quasi-modes exhibiting sharp resonances whose linewidths are dictated by the mirrors' properties. The recycling of photons in the confined volume allows, e.g., for enhanced near-resonant interactions with quantum emitters when the frequency of the quasi-mode is close to the emitters' transition frequency~\cite{Haroche2006}, or for strongly frequency-dependent radiation pressure forces acting on the mirrors~\cite{Aspelmeyer2014}. The quasi-mode approximation allows for the ad-hoc derivation and treatment of simple, convenient light-matter interaction Hamiltonians, such as the Jaynes-Cummings Hamiltonian or the dispersive radiation pressure Hamiltonian, which are widely used in quantum optics.

\begin{figure}[t]
\includegraphics[width=0.99\columnwidth]{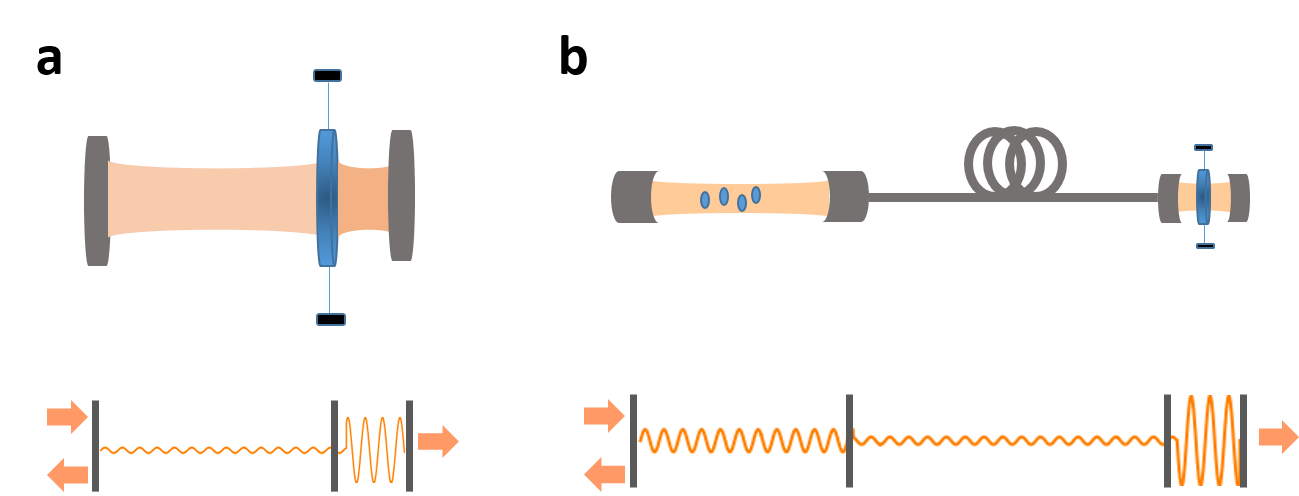}
\caption{\emph{Multi-element optical structures} -- (a) Illustration of a three-element 'asymmetric' configuration modelling, e.g., membrane-based optomechanics (top) and its idealized one-dimensional respresentation suggesting the confinement of light within the smaller subcavity (bottom). (b) Four-element configuration modelling, e.g., a hybrid atom-optomechanics modular architecture in which a fiber-cavity optomechanical system is coupled via an optical link to another fiber-cavity system containing an ensemble of quantum emitters (top) and the corresponding one-dimensional four-element representation (bottom).} \label{fig:fig1}
\end{figure}

A variety of modern experiments involve, however, optical systems involving more than two elements. For example, a three-element resonator geometry in which two fixed mirrors surround a vibrating membrane is widely used within the field of optomechanics either in traditional Fabry-Perot~\cite{Thompson2008,Wilson2009,Karuza2012,Kemiktarak2012NJP,Purdy2013} or fiber-based~\cite{Flowers2012,Shkarin2014} setups (Fig.~\ref{fig:fig1}a). Modular setups involving coupled cavities are also extensively used in cavity QED networks with quantum emitters~\cite{Cirac1997,Pellizzari1997,VanEnk1999,Serafini2006,Kimble2008,Hartmann2008,Lepert2011,Ritter2012,Kurizki2015}, optomechanics with multiple resonators~\cite{Bhattacharya2008multiple,Hartmann2008,Stannigel2010,Heinrich2011,Massel2012,Xuereb2012,Zhang2012,Stannigel2012,Bagheri2013,Zhang2015,Fang2016,Spethmann2016,Nair2016,Nair2017}, or in hybrid optomechanics~\cite{Treutlein2014,Wallquist2009}, where single~\cite{Tian2004,Favero2008,Singh2008,Hammerer2009strong,Restrepo2014} or ensembles~\cite{Meiser2006,Treutlein2007,Genes2008emergence,Ian2008,Genes2009micromechanical,Hammerer2009establishing,Hunger2010resonant,Paternostro2010,Camerer2011,Genes2011,Vogell2013,Dantan2014,Vogell2015,Molmer2016} of quantum emitters are interfaced with mechanical resonators (Fig.~\ref{fig:fig1}b).

The presence of multiple optical elements, not necessarily all highly reflecting, renders the description of the confined field modes more intricate. A number of models for coupled cavity arrays therefore use an {\it ad hoc} combination of independent Hamiltonians in the various subcavities and mutual couplings (e.g. tunneling). However, while one can usually find reasonable regimes in which this treatment is valid, this approach potentially ignores interference effects between the multiple elements which can make the properties of the individual subcavities dependent on the global properties of the compound system (effective decay rate, effective modevolume, etc.). For instance, a rigorous quantization of the mode functions of a four-mirror cavity~\cite{Ley1987,VanEnk1999} can be required to identify the relevant interaction parameters for a pair of fiber-coupled cavities~\cite{Cirac1997,Pellizzari1997,Serafini2006} or the regimes in which tunneling between distinct cavity modes becomes important~\cite{Dombi2013}.

In this paper we aim at providing a roadmap, based on first principles, for rigorously deriving the interaction coefficients to be subsequently used in the simplified quantum optics Hamiltonians describing the dynamics of multi-element optical resonators. Our approach is based on the well-established one-dimensional transfer matrix formalism derived from the decomposition of the Helmholtz equation in equations for independent plane wave components~\cite{Deutsch1995}. This approach allows for constructing the field distribution between the resonators, computing the resonances of the compound system and rigorously identifing regimes in which interactions with a single extended cavity mode are justified and regimes in which tunneling between distinct modes (see e.g.~\cite{VanEnk1999,Dombi2013}) has to be taken into account. In the single mode regime we provide a recipe for deriving the effective cavity field decay rate, the Jaynes-Cummings coherent coupling strengths of the confined fields with two-level quantum emitters or the cavity optomechanical couplings with movable reflectors. We analyze in particular resonators comprised of three and four elements, which are relevant for a wide range of (hybrid) optomechanical setups~\cite{Thompson2008,Wilson2009,Karuza2012,Kemiktarak2012NJP,Purdy2013,Flowers2012,Shkarin2014,Hammerer2009strong,Hammerer2009establishing,Hunger2010resonant,Camerer2011,Genes2011,Vogell2013,Vogell2015} or systems of coupled (fiber-) cavities with atoms, NV-centers in diamond, quantum dots, molecules, etc.~\cite{Kimble2008,Hartmann2008,Kurizki2015,Takahashi2013,Steiner2013,Brandstatter2013,Colombe2007,Becher2013,MiguelSanchez2013,Kelkar2015} and identify situations in which confining the field in specific subcavities leads to enhanced cavity QED and/or optomechanical interactions.

The paper is organized as follows: we introduce in Sec.~\ref{sec:two} the transfer matrix model in a canonical two-mirror resonator situation. In Sec.~\ref{sec:three} we proceed with a numerical analysis of a three-mirror system and provide a generic analytical method to compute the resonances of the compound system, their linewidths and the field distribution in the subcavities. As an example, we focus on the situation in which the middle reflector is movable and subjected to radiation pressure; we first analyze the single mode approximation validity and, second, show that enhanced optomechanical coupling can be achieved in a strongly 'asymmetric' cavity configuration. In Sec.~\ref{sec:four} we extend the analysis to a four-mirror configuration and show that a simple analytical treatment adequately describes the system in the single mode regime and provides a simple path for the derivation of both the Jaynes-Cummings and dispersive radiation pressure optomechanical coupling strengths. We show in particular that, in the case of two short cavities separated by a long one and for specific resonances of the compound system, the interaction with a single extended cavity mode for which the field is essentially confined to the short cavities allows for enhanced light-matter interactions inside them.

\section{Two-mirror resonator}\label{sec:two}

\subsection{Transfer matrix model}

We consider a parallel Fabry-Perot arrangement and make use of a standard one-dimensional transfer matrix formalism~\cite{Deutsch1995} to describe the static properties of the system and to subsequently infer its dynamics. Each optical element is described by a transfer matrix $M$ which relates the forward- and backward-propagating waves on each side
\begin{equation}
\biggl(\begin{array}{c}A\\B\end{array}\biggr)=M\biggl(\begin{array}{c}C\\D\end{array}\biggr)=\biggl[\begin{array}{cc}m_{1,1} & m_{1,2}\\ m_{2,1} & m_{2,2}\end{array}\biggr]\biggl(\begin{array}{c}C\\D\end{array}\biggr),
\end{equation}
with $A$ and $C$ ($B$ and $D$) the amplitudes of the backward-propagating (forward-propagating) waves. The free-space propagation of a monochromatic field with wavenumber $k$ over the distance $L$ between the two mirrors is described by the matrix
\begin{equation}
M_\text{fs}(\theta)=\biggl[\begin{array}{cc}e^{i\theta} & 0\\0 & e^{-i\theta}\end{array}\biggr],
\end{equation}
with $\theta=kL$. The mirrors are modelled by infinitely thin, absorptionless dielectric slabs characterized by their polarizability $\zeta$ and described by the matrix
\begin{equation}
M(\zeta)=\biggl[\begin{array}{cc}1+i\zeta & i\zeta\\-i\zeta & 1-i\zeta\end{array}\biggr].
\end{equation}
The left and right mirrors are assumed to have polarizabilities $\zeta$ and $\zeta'$, respectively. To simplify, we neglect the variation of the polarizabilities with frequency. The total transfer matrix of the system can then be written as
\begin{equation}
M_{\textrm{II}}=M(\zeta)M_\text{fs}(\theta)M(\zeta')\,.
\end{equation}
The transmission of the system is given by $\mathcal{T}=1/|(M_{\textrm{II}})_{2,2}|^2=1/D(\theta)$, where \begin{equation}
D(\theta)=\left|(1-i\zeta)(1-i\zeta')e^{-i\theta}+\zeta\zeta'e^{i\theta}\right|^2\,.
\end{equation}
Resonances are found when the denominator is minimum, which occurs when $D'(\theta_0)=0$, with $\theta_0$ given by
\begin{equation}
\tan(2\theta_0)=-\frac{\zeta+\zeta'}{1-\zeta\zeta'}\,.
\label{eq:theta0}
\end{equation}
The linewidth of a well-resolved resonance can be found by expanding $D(\theta_0+\delta\theta)$ around $\theta_0$ at second order in $\delta k$, where $\delta\theta=L\delta k$:
\begin{align}
D(\theta_0+\delta\theta)=D(\theta_0)\left[1+\frac{D''(\theta_0)}{2D(\theta_0)}\left(L\delta k\right)^2\right]\,,
\end{align}
which yields a HWHM (half-width at half maximum) in angular frequency $\omega=ck$ given by
\begin{equation}
\kappa=\frac{c}{L}\sqrt{\frac{2D(\theta_0)}{D''(\theta_0)}}\,,
\end{equation}
where $c$ is the speed of light in vacuum. In the two-mirror situation considered here, one obtains
\begin{equation}
\kappa=\frac{c}{2L}\frac{\sqrt{(1+\zeta^2)(1+\zeta'^2)}-\zeta\zeta'}{\sqrt{\zeta\zeta'}\sqrt[4]{(1+\zeta^2)(1+\zeta'^2)}}\,.
\end{equation} When the mirror reflectivities are close to one ($\zeta,\zeta'\gg 1$), one retrieves that the linewidth is given by the product of the cavity free-spectral range by the sum of the transmission losses of the mirrors, i.e.,
\begin{equation}
\kappa\simeq\frac{c}{4L}\left(\frac{1}{\zeta^2}+\frac{1}{\zeta'^2}\right)\,.
\end{equation}
Denoting the field amplitude inside the cavity as $E(x)=C_+e^{ikx}+C_-e^{-ikx}$ and assuming for instance a right-propagating monochromatic plane wave, incident from the left, its mean intensity $|E|^2=|C_+|^2+|C_-|^2$ can be computed straightforwardly by application of the transfer matrix formalism, up to a normalization constant. In the two-mirror situation, one gets
\begin{equation}
|E|^2=\frac{1+2\zeta'^2}{(\sqrt{(1+\zeta^2)(1+\zeta'^2)}-\zeta\zeta')^2}\,.
\end{equation}

\subsection{Dispersive optomechanics}
For applications in the context of optomechanics let us assume that the right mirror with polarizability $\zeta'$ is movable and harmonically bound around its equilibrium position. Its optomechanical coupling can be derived by computing the shift in the cavity field resonance frequency $\delta\omega$ when the mirror is slightly displaced around its rest position by $\delta x$. When the frequency shift $\delta\omega=c\delta k$ is linear in the displacement, the linear dispersive optomechanical coupling (single photon-single phonon coupling rate) is given by $G=c|{\delta k}/{\delta x}|$ and can be used to construct the reduced single-mode model radiation pressure Hamiltonian
\begin{equation}
H_{\textrm{OM}}=\hbar G\hat{a}^{\dagger}\hat{a}\hat{x}\,,
\end{equation}
where $\hat{a}$ and $\hat{a}^{\dagger}$ are the annihilation and creation operators for the field mode associated with the resonance considered and $\hat{x}$ is the displacement operator for the vibrational mode~\cite{Aspelmeyer2014}.

Assuming again a well-resolved cavity resonance, the optomechanical coupling of the movable mirror is calculated via a first order expansion of $D(\theta)$ around the resonance, namely, by minimizing $D(\theta_0+\delta\theta)$, where $\delta\theta=L\delta k+k\delta x$. For small displacements of the mirror one indeed finds a linear scaling of $\delta k$ with $\delta x$, resulting in the usual optomechanical coupling~\cite{Aspelmeyer2014}
\begin{equation}
G=\frac{ck}{L}.
\end{equation}
A relevant figure of merit in many optomechanical applications~\cite{Aspelmeyer2014} is the optomechanical cooperativity, $C^{\textrm{OM}}\varpropto G^2/\kappa$, which scales as $1/L$.

\subsection{Cavity QED}

For applications in the context of cavity QED let us assume that a two-level system is positioned in one of the subcavities and couples to the field via a Jaynes-Cummings interaction. Let us also make the assumptions that the cavity is large enough and sustains a large enough solid angle, so that the cavity field modefunction is not modified by the presence of the emitter and the emitter's coupling to electromagnetic modes other than that of the cavity via spontaneous emission (at rate $\gamma$) is not modified. We further assume that the cavity resonance is well-resolved and that its frequency corresponds to the transition frequency $\omega$ between the two levels. We thus consider a Jaynes-Cummings interaction (in the rotating wave approximation) of the form
\begin{equation}
H_{\textrm{JC}}=\hbar g(\hat{a}\hat{\sigma}_++\hat{a}^{\dagger}\hat{\sigma}_-)\,,
\end{equation}
where $\hat{a}$ and $\hat{a}^{\dagger}$ are the annihilation and creation operators for the mode considered and $\hat{\sigma}_+$ and $\hat{\sigma}_-$ the quantum emitter's raising and lowering operators. We wish to compute the maximal coupling strength of the quantum emitter with the field, $g=d\mathcal{E}_0/\hbar$, where $d$ is the strength of the dipole element and $\mathcal{E}_0$ the rms amplitude of the vacuum field for the mode considered. The relevant mode volume is defined via the energy relation for the quantized vacuum field $E_{\textrm{vac}}$
\begin{equation}\label{eq:energy}
\int |E_{\textrm{vac}}|^2dV =
\frac{1}{2}\varepsilon_0AL\mathcal{E}_{0,L}^2=\frac{1}{2}\hbar\omega\,,
\end{equation}
where $A$ is the cavity mode transverse cross-section and $\varepsilon_0$ the vacuum permittivity. This relation yields the standard expression for $g$~\cite{Haroche2006}
\begin{equation}
g=\frac{d}{\hbar}\sqrt{\frac{\hbar\omega}{\varepsilon_0AL}}.
\end{equation}
A relevant figure of merit for many cavity QED interactions~\cite{Kimble2008} is the (``Jaynes-Cummings") cooperativity, $C= g^2/\kappa \gamma$, which, in the idealized 1D geometry, is independent of the length.

As we will illustrate in Sec.~\ref{sec:four} in a more complex scenario, combining the energy relation from Eq.~(\ref{eq:energy}) and the knowledge of the field distribution computed from the transfer matrix allows for defining the relevant effective mode volume and coupling strength between the emitter and the field, which can subsequently be used in a reduced single-mode cavity QED description of the dynamics. The transfer matrix approach thus conveniently provides, at least in the case of well-resolved cavity resonances, all the necessary ingredients required to build standard single-mode quantum optics models for treating the dynamics of two-level systems or movable elements interacting with the field of the resonator.

\section{Three-element resonator}\label{sec:three}

\begin{figure}[t]
\includegraphics[width=0.59\columnwidth]{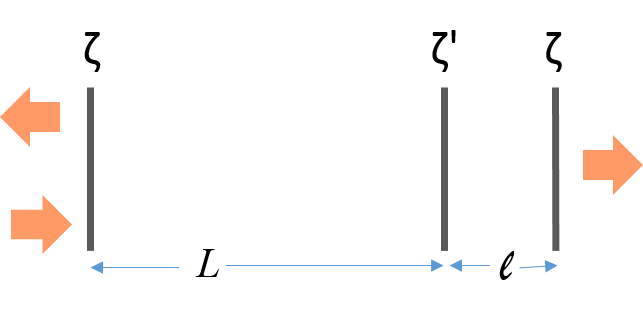}
\caption{\emph{Three-element resonator} -- Three optical elements with polarizabilities $\zeta$, $\zeta'$ and $\zeta$ are positioned at $0$, $L$ and $L+l$. The system's optical properties are computed by assuming a plane wave (with unit amplitude) incident from the right, which results in reflected and transmitted components.} \label{fig:fig2}
\end{figure}

We consider the three-mirror arrangement depicted in Fig.~\ref{fig:fig2}, where two outer mirrors with polarizability $\zeta$ and a middle mirror with polarizability $\zeta'$ form left- and right-subcavities with lengths $L$ and $l$, respectively. The total transfer matrix of the system can then be written as
\begin{equation}
M_{\textrm{III}}=M(\zeta)M_\text{fs}(\theta)M(\zeta')M_\text{fs}(\phi)M(\zeta)\,,
\end{equation}
where the accumulated phases during free-propagation are $\theta=kL$ and $\phi=kl$. The transmission of the compound system is given by
\begin{equation}
\mathcal{T}=\frac{1}{D(\theta,\phi)},
\end{equation}
where $D(\theta,\phi)=|(M_{\textrm{III}})_{2,2}|^2$.  The transmission resonances and their respective linewidths can be found as previously by analyzing the minima of $D(\theta,\phi)$. We first start by discussing the behavior of this compound system on a numerical example before interpreting the results analytically.

\begin{figure}[t]
\includegraphics[width=\columnwidth]{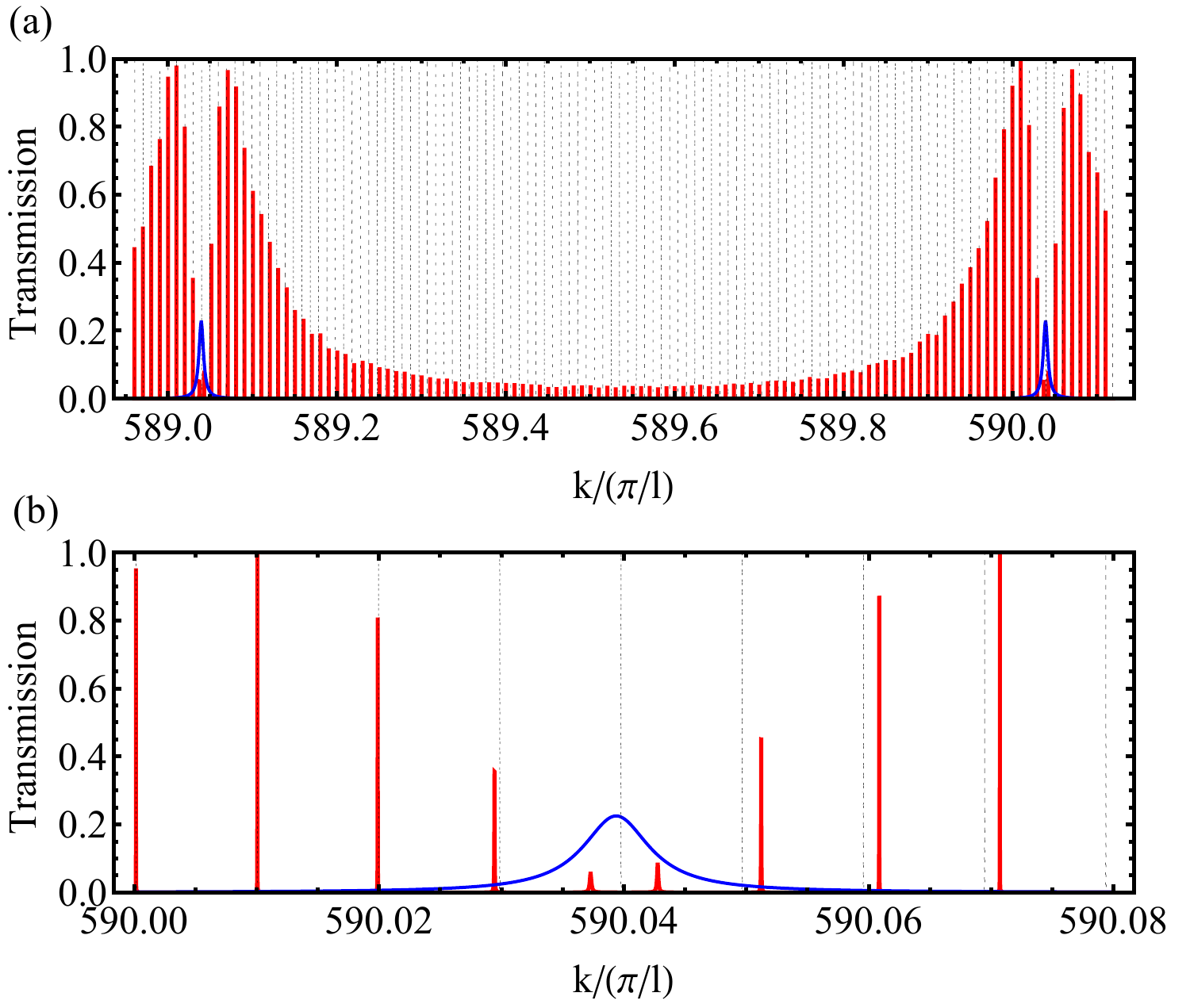}
\caption{\emph{Resonances of a three-element resonator: single extended mode regime} -- Transmission of a three-mirror 'asymmetric' cavity with $\zeta=20$, $\zeta'=5$, $L=100\pi$, $l=\pi$ over a range corresponding to a free-spectral range of the cavity with length $l$ (a) and over a narrower range around a common resonance (b). For comparison, the spectra of a symmetric two-mirror cavity with length $L+l$ and $\zeta=20$ as well as that of a short two-mirror cavity with length $l$ and $\zeta=20$ and $\zeta'=5$ are shown in black and blue, respectively.}
\label{fig:example}
\end{figure}

As a first example, Fig.~\ref{fig:example} shows the transmission of an 'asymmetric' three-mirror cavity, consisting of two highly-reflecting mirrors ($\zeta=20$) and a poorly-reflecting mirror ($\zeta'=5$) in between them, but positioned close to one of the end-mirrors ($l/L=1/100$). The transmission spectrum of the compound cavity is a complex one, with resonances showing varying transmission levels depending on the interferences between the subcavity fields. The majority of the resonances (red) are observed for frequencies which are multiple of the free-spectral range of the large cavity with length $L+l$, and the level of transmission increases as the frequency gets close to a resonance frequency of the short cavity with length $l$ (blue), since the electric field intensity in the large subcavity increases to the level of a symmetric cavity of length $L$. However, for 'common resonances', i.e., the short and long cavity resonances overlap, the left and right cavity modes hybridize and light can be essentially confined in the short subcavity with length $l$. Consequently, the resonance peaks there exhibit reduced transmission and a broadened linewidth, as will be discussed in more detail later on.

\begin{figure}[t]
\includegraphics[width=\columnwidth]{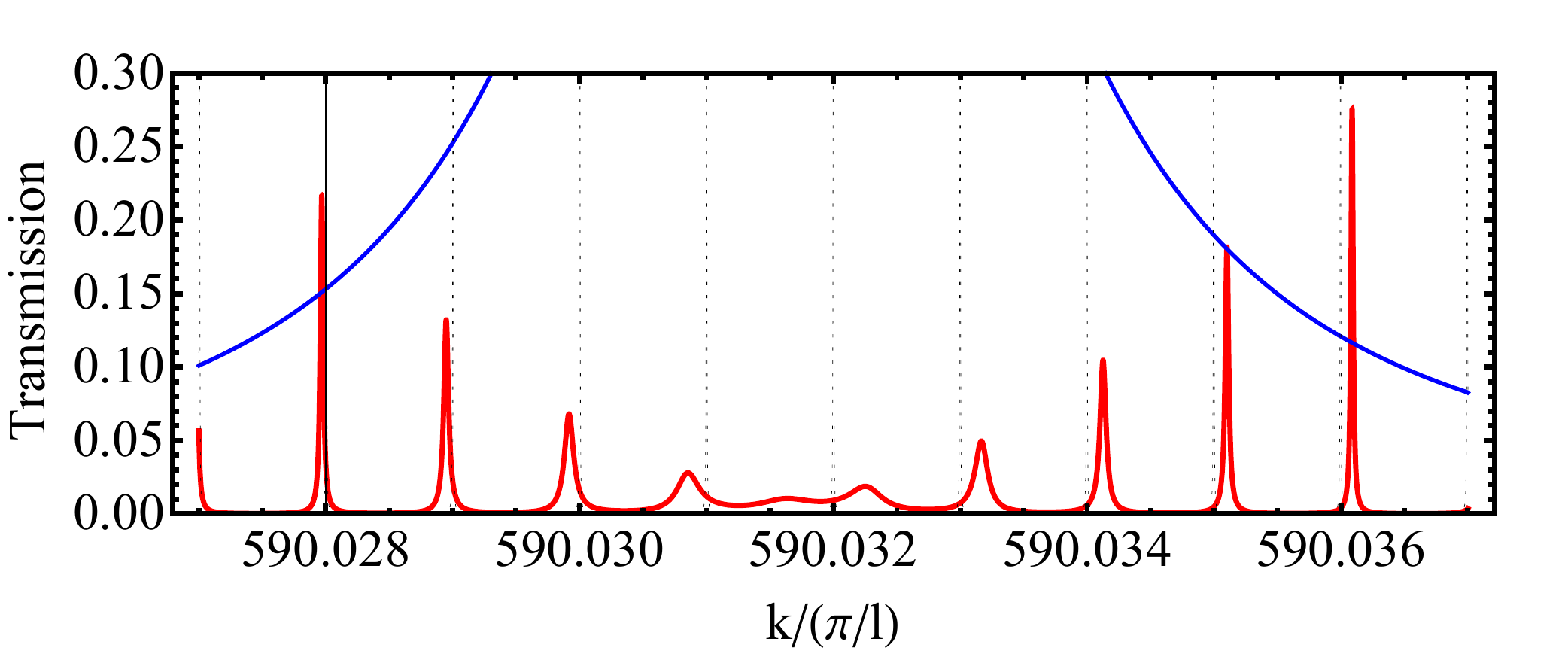}
\caption{\emph{Resonances of a three-element resonator: tunneling regime} -- Transmission of a three-mirror 'asymmetric' cavity with $\zeta=\zeta'=10$, $L=1000\pi$, $l=\pi$ around a common resonance. For comparison, the spectra of a symmetric two-mirror cavity with length $L+l$ and $\zeta=10$ as well as that of a short, symmetric two-mirror cavity with length $l$ and $\zeta=10$ are shown in black and blue, respectively.}
\label{fig:example2}
\end{figure}

A second example is shown in Fig.~\ref{fig:example2} in the case of an even more asymmetric three-mirror cavity ($\l/L=1000$) and for identical highly-reflecting mirrors ($\zeta=\zeta'=10$). The compound system transmission spectrum around a 'common resonance' clearly shows partially overlapping resonances, as the width of the hybridized modes become comparable to the spacing between them. In this strong tunneling regime, whose emergence will also be discussed later, the single-mode description is not applicable.

\subsection{Numerical results}

\begin{figure}[t]
\centering
\includegraphics[width=0.98\columnwidth]{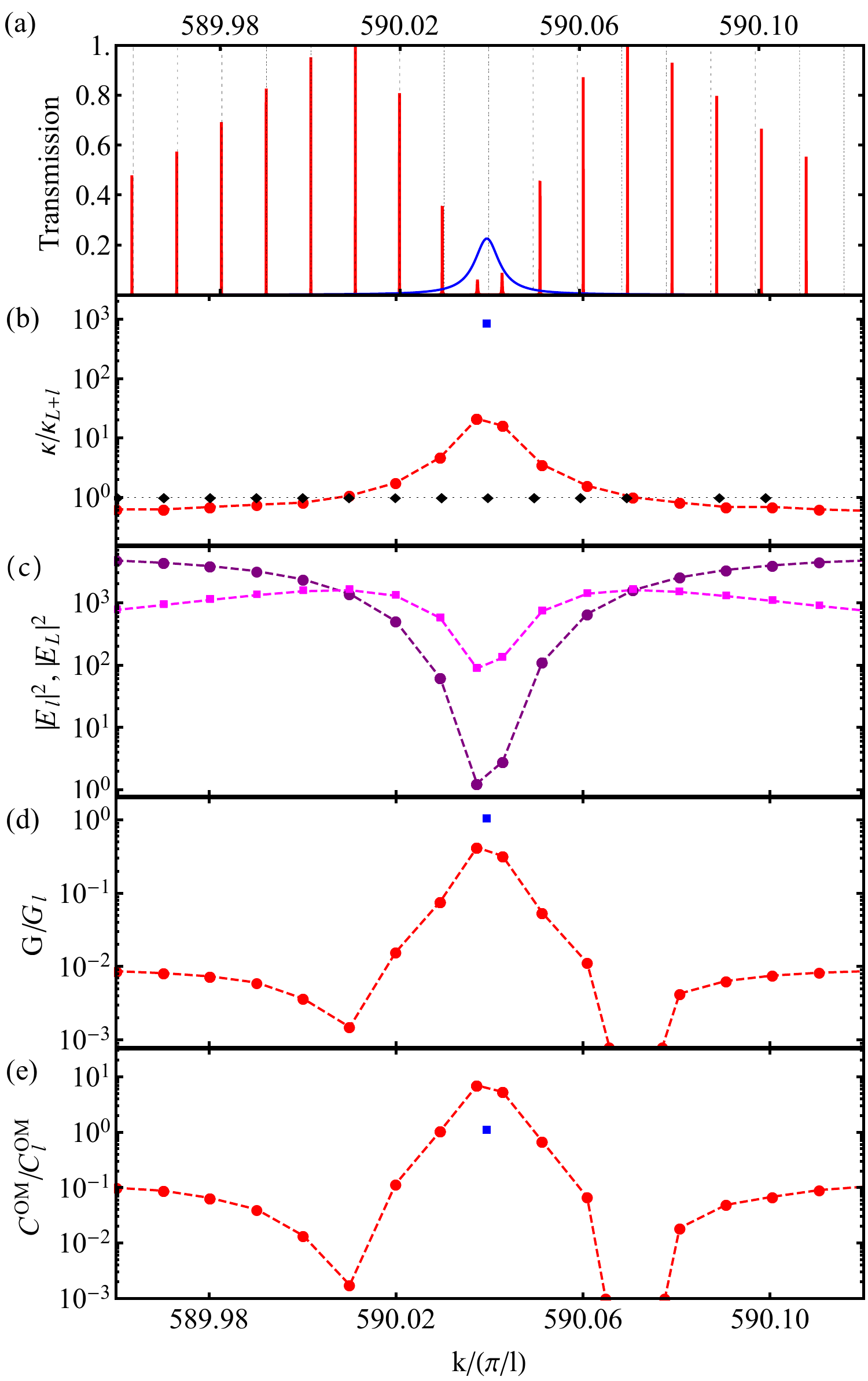}\\
\caption{(a) Transmission spectra (red) as a function of $k/(\pi/l)$ around a 'common' resonance for a compound cavity with $\zeta=20$, $\zeta'=5$, $L=100\pi$, $l=\pi$. The black and blue curves show the spectrum of a cavity with no middle mirror and total length $L+l$ and of the right subcavity with length $l$, respectively. (b) Half-width-half-maximum (HWHM) (red dots) of the resonances shown in (a), normalized to that of a cavity with no middle mirror and length $L+l$ (blue square). (c) Mean field intensities in the left (purple circles) and right (magenta squares) subcavities. The incident field amplitude is normalized to unity. (d) Linear optomechanical coupling (red dots) associated with each resonance of the compound cavity, normalized to that of the right subcavity with length $l$ (blue square). (e) Corresponding optomechanical cooperativity (red dots), normalized to that of the right subcavity with length $l$ (blue square). The dashed lines are merely a guide for the eye.} \label{fig:fig3}
\end{figure}

We first numerically investigate the optomechanical properties of such a strongly asymmetric cavity ($L/l=100$) with a middle mirror whose reflectivity is somewhat less than that of the outer mirrors ($\zeta'=5$, $\zeta=20$). We assume that the middle reflector is harmonically bound around its equilibrium position with oscillation amplitudes well into the Lamb-Dicke regime, as is customary for typical optomechanical resonators~\cite{Aspelmeyer2014}.  This situation could thus represent an alternative -- 'membrane-at-the-end' -- geometry to the widely used experimental "membrane-in-the-middle" configuration~\cite{Thompson2008}.

Figure~\ref{fig:fig3}a shows the compound cavity transmission spectrum (red) for $k$ vectors close to a resonance of the short, right subcavity resonance (blue) given by  $k_l\simeq 590\pi/l$. For comparison the resonance spectrum of a cavity with the same total geometrical length $L+l$, but no middle mirror ($\zeta'=0$), is also shown (black). As discussed previously, the compound cavity transmission is observed to sharply drop for $k$ vectors very close to the short cavity resonance at $k_l$. As can be seen in Fig.~\ref{fig:fig3}b the linewidth concomitantly increases as a consequence of the hybridization of the field modes of both subcavities. One also observes a strong increase in the optomechanical coupling strength (Fig.~\ref{fig:fig3}c), which becomes comparable to the level of coupling that would be obtained in the short right subcavity alone. Figure~\ref{fig:fig3}d remarkably shows that the optomechanical cooperativity can even increase above the short cavity value.

These results can be interpreted by computing the field intensities in the left and right subcavities. Figure~\ref{fig:fig3}e shows the variation of the mean intensities of the subcavity fields for various resonances of the compound cavity. Around the 'common' resonance, most of the field accumulates in the short right subcavity; this results in a shorter effective length of the compound cavity and thereby an increased linewidth, but also in a stronger radiation pressure force on the middle mirror which leads to stronger optomechanical coupling and cooperativity.

\subsection{Analytical results}

In order to derive analytical expressions and provide a generic interpretation of these results, we first note that resonances occur in the left subcavity  when $\theta=\theta_0$, with $\theta_0$ defined by~(\ref{eq:theta0}) and in the right subcavity when $\phi=\phi_0$, with $\phi_0$ defined when substituting $\theta_0$ by $\phi_0$ in~(\ref{eq:theta0}). Let us fix the cavity lengths and polarizabilities and assume that there exists a wavenumber $k_0$ such that the following resonance conditions in both cavities
\begin{subequations}
\begin{align}
\label{eq:res3a}\theta_0&=-\frac{1}{2}\arctan\left(\frac{\zeta+\zeta'}{1-\zeta\zeta'}\right)+\frac{\pi}{2},\\
\label{eq:res3b}\phi_0&=-\frac{1}{2}\arctan\left(\frac{\zeta+\zeta'}{1-\zeta\zeta'}\right),
\end{align}
\label{eq:res3}
\end{subequations}
can be simultaneously satisfied. This choice of phases corresponds to the right matching conditions at the middle mirror for constructive field buildup in the compound cavity. One can then compute the linewidth by expanding $D(\theta_0+\delta\theta,\phi_0+\delta\phi)$, where $\delta\theta=L\delta k$ and $\delta\phi=l\delta k$, at second order in $\delta k$. An exact expression, valid for arbitrary $\zeta$ and $\zeta'$, of the obtained linewidth (HWHM) for such a common resonance is given in the Appendix. In the regime $\zeta\gg 1$, one finds a simple expression for the common resonance linewidth
\begin{equation}
\kappa\simeq\frac{c}{2\zeta^2}\frac{\sqrt{1+\zeta'^2}}{L(\sqrt{1+\zeta'^2}-\zeta')+l(\sqrt{1+\zeta'^2}+\zeta')}=\frac{c}{2\zeta^2L_{\textrm{eff}}}\,,
\end{equation}
where $L_{\textrm{eff}}$ represents the effective length of the compound cavity
\begin{equation}
L_{\textrm{eff}}=L(1-r')+l(1+r')
\end{equation}
and $r'=\zeta'/\sqrt{1+\zeta'^2}$ the absolute value of the middle mirror reflectivity.

First, let us note that, when $\zeta'=0$ or when the movable mirror is at the geometric center of the cavity (``membrane-in-the-middle", $L=l$), the effective length corresponds to the total geometrical length $l+L$. In contrast, when $\zeta'\gg 1$, the effective cavity length becomes (twice) the length of the right subcavity. In other words, in an asymmetric cavity ($L\gg l$), the linewidth for the common resonance is substantially broadened as compared to the linewidth of the long cavity resonance, and essentially becomes comparable to (half) the small cavity linewidth. As aforementioned, this results from the fact that the field amplitude is greater in the small cavity than in the large cavity. The factor two is a consequence of the fact that, in the regime $\zeta\gg 1$, the left subcavity becomes equivalent to a perfectly reflecting mirror for the right subcavity which then sees its transmission losses divided by two.

When $\zeta\gg 1$, one can indeed show that the mean field intensities in the left and right subcavities are
\begin{equation}
|E_L|^2\simeq2\zeta^2\frac{(\sqrt{1+\zeta'^2}-\zeta')^2}{1+\zeta'^2}\,
\end{equation} and
\begin{equation}
|E_l|^2\simeq\frac{2\zeta^2}{1+\zeta'^2}\,,
\end{equation}
respectively. Their ratio
\begin{equation}
|E_L/E_l|^2=(\sqrt{1+\zeta'^2}-\zeta')^2
\end{equation}
clearly shows the increase (decrease) of the light intensity in the small (large) cavity as $\zeta'$ increases.

\begin{figure}[t]
\centering
\includegraphics[width=0.85\columnwidth]{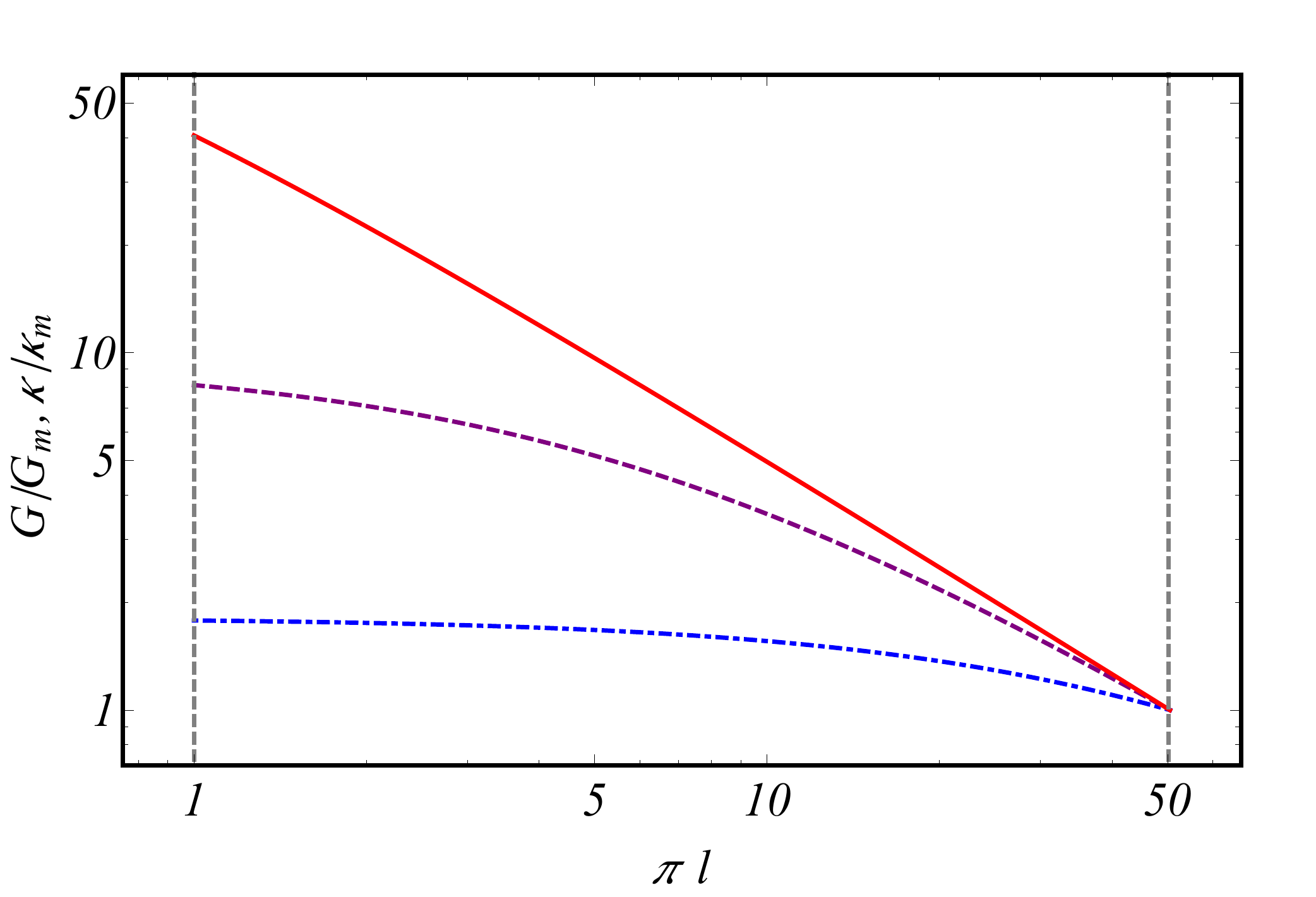}
\caption{Variation of $G$ and $\kappa$, normalized to their respective value in the mirror-in-the-middle situation, as a function of the length $l$ of the right subcavity, when the total geometrical length $L+l$ of the three-mirror cavity is kept constant and both resonance conditions~(\ref{eq:res3a},\ref{eq:res3b}) are simultaneously satisfied. Parameters: $\zeta=20$, $L+l=101\pi$ and $\zeta'=0.5$ (dot-dashed), 2 (dashed) and 10 (plain). The vertical dashed lines indicate the asymmetric (left) and symmetric (right) cavity configurations.} \label{fig:fig4}
\end{figure}

Let us point out that the above analysis is only valid if the common resonance considered is well-resolved, i.e., does not significantly overlap with nearby resonances. This condition requires that the effective linewidth is less than the free spectral range of the large cavity $c/2(L+l)$, which determines the spacing between adjacent resonances. In the regime $\zeta,\zeta'\gg 1$ and for $L\gg l$, this non-overlapping condition reads
\begin{equation}
\sqrt{\frac{2l}{L}}>\frac{\zeta'}{\zeta^2}\,.
\end{equation} When the non-overlapping condition is not met, significant tunneling between the modes can occur, as illustrated in Fig.~\ref{fig:example2}, and the single-mode description is no longer valid (see also~\cite{Genes2013} for a discussion of coalescing symmetric cavities).

Assuming that the common resonance is well-resolved, the optomechanical coupling of the movable middle mirror can also be calculated by expanding $D(\phi,\theta)$ at first order around the resonance, i.e., by minimizing $D(\phi_0+\delta\phi,\theta_0+\delta\theta)$, where $\delta\theta=L\delta k+k\delta x$ and $\delta\phi=l\delta k-k\delta x$. For small displacements of the mirror one finds a linear scaling of $\delta k$ with $\delta x$. An exact expression for the linear optomechanical coupling, valid for any values of $\zeta$ and $\zeta'$, is given in the Appendix. In the regime $\zeta\gg 1$ (highly-reflecting outer mirrors), the coupling takes on the simple expression
\begin{equation}
G\simeq\frac{2ck\zeta'}{(\sqrt{1+\zeta'^2}-\zeta')L+(\sqrt{1+\zeta'^2}+\zeta')l}=\frac{2ck}{L_{\textrm{eff}}}r'\,.
\end{equation}
When the movable mirror is in the middle ($L=l$) one retrieves that $G=(2ck/L)r'\equiv G_m$~\cite{Thompson2008}. In contrast, when the cavity is asymmetric ($L\gg l$) and $\zeta'\gg 1$, one has
\begin{equation}
\frac{G}{G_m}\simeq\frac{1}{1-r'+(l/L)(1+r')}\,,
\end{equation}
showing that $G$ increases substantially when $r'$ increases, and tends towards $ck/l$, which is (half) the value of the optomechanical coupling of a perfectly reflecting movable mirror in a small cavity with length $l$.

Given that both $\kappa$ and $G$ scale with the inverse of the effective cavity length, it is clear that the optomechanical cooperativity,  proportional to $G^2/\kappa$, increases when going from a symmetric to an asymmetric cavity configuration. This is illustrated in Fig.~\ref{fig:fig4}, which shows the variation of the linewidth and the optomechanical coupling, normalized to their respective values in the mirror-in-the-middle configuration, as a function of the length of the right subcavity, for a constant total geometrical length $L+l$ and when both resonance conditions are simultaneously satisfied. While the increase in $G$ or $\kappa$ remains moderate in the case of a low-reflectivity middle mirror~\cite{Thompson2008}, it becomes quite substantial for a high-reflectivity mirror~\cite{Kemiktarak2012NJP,Reinhardt2016,Norte2016,Chen2016} whose reflectivity is large enough for the effective length of the short cavity $4\zeta'l$ to be comparable or larger than the large subcavity length $L$. We note that exactly enforcing the 'common resonance' condition is not a necessary condition (the arbitrary case of Fig.~\ref{fig:fig3} is an example), which means that a precise control of the position of the membrane at the end is not required experimentally. By adjusting the laser frequency to be close to a resonance frequency of the short cavity, one can always find a compound resonance exhibiting enhanced OM coupling. In that sense, the experimental constraints regarding the positioning of the 'membrane-at-the-end' are similar to those of the 'membrane-in-the-middle' geometry. We also point out that such a 'membrane-at-the-end' geometry has already been implemented in~\cite{Purdy2012,Purdy2013}, although with lower asymmetry level ($l/L\simeq 0.18$) and lower membrane reflectivity ($\zeta'\simeq 0.3$). The factor $\sim 50$ enhancement shown in Fig.~\ref{fig:fig4} could be obtained with, e.g., a $\sim 1$ cm-long close-to-hemifocal cavity, similar to that used in~\cite{Purdy2012,Purdy2013}, and a highly-reflecting membrane ($>99\%$, as used in~\cite{Kemiktarak2012NJP,Norte2016,Reinhardt2016}), positioned $\sim 100$ microns from the plane cavity mirror. We also checked numerically that taking into account a realistic membrane thickness, as in Ref.~\cite{Nair2016}, or absorption, as in Ref.~\cite{Xuereb2013}, does not appreciably change the results.

\section{Four-element resonator}\label{sec:four}

\begin{figure}[h]
\centering
\includegraphics[width=0.95\columnwidth]{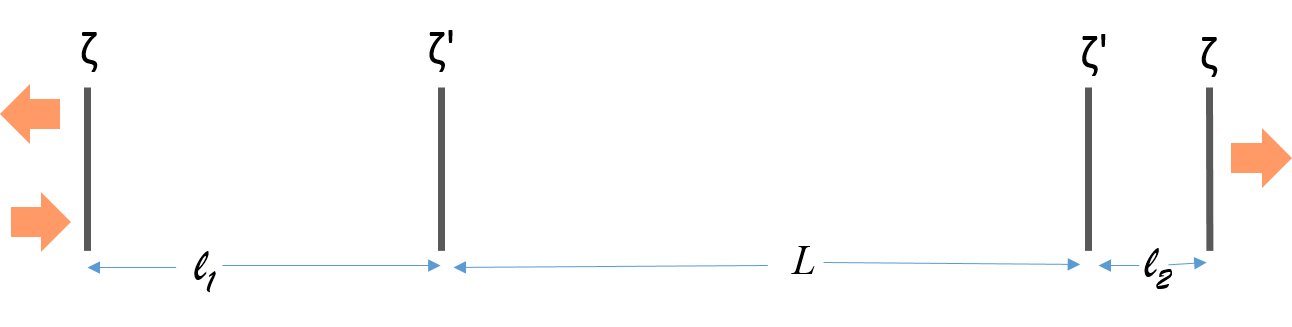}
\caption{\textit{Four-element configuration}- Four optical elements with polarizability $\zeta$, $\zeta'$, $\zeta'$ and $\zeta$ are positioned at $0$, $l_1$, $L+l_1$ and $L+l_1+l_2$.} \label{fig:fig5}
\end{figure}

We now consider the generic four-mirror configuration depicted in Fig.~\ref{fig:fig5} in which four mirrors are positioned at $x=0$, $l_1$, $l_1+L$ and $l_1+L+l_2$ and have polarizabilities $\zeta$, $\zeta'$, $\zeta'$ and $\zeta$, respectively. Setting $\theta=kL$ and $\phi_i=kl_i$ ($i=1,2$), the transfer matrix of the four-mirror system is given by the product
\begin{equation}
M(\zeta)M_{\textrm{fs}}(\phi_1)M(\zeta')M_{\textrm{fs}}(\theta)M(\zeta')M_{\textrm{fs}}(\phi_2)M(\zeta).
\end{equation}
While the optical and optomechanical properties of the system can be numerically computed in any arbitrary configuration using the previously described method, we focus in the next section on the symmetrical configuration $l_1=l_2=l$, which allows for tractable analytical derivations. The situation of left/right subcavities with different lengths, in particular $l_1 \gg l_2$, will be discussed in Sec.~\ref{sec:modular} in the context of hybrid modular optomechanics.

\subsection{Transmission spectrum}

In this symmetrical situation we assume that there exists a wave-vector $k$ such that the resonance conditions
\begin{subequations}
\begin{align}
\label{eq:res4a}\theta_0&=-\frac{1}{2}\arctan\left(\frac{2\zeta'}{1-\zeta'^2}\right),\\
\label{eq:res4b}\phi_0&=-\frac{1}{2}\arctan\left(\frac{\zeta+\zeta'}{1-\zeta\zeta'}\right),
\end{align}
\label{eq:res4}
\end{subequations}
are simultaneously satisfied. This choice of phases corresponds again to constructive buildup of the field at both middle mirrors. When $\zeta\gg 1$, the linewidth is approximately given by (the general expression valid for arbitrary $\zeta$ is given in the Appendix)
\begin{equation}
\label{eq:kappa4}
\kappa\simeq\frac{c}{2\zeta^2}\frac{1}{2l+L(\sqrt{1+\zeta'^2}-\zeta')^2}=\frac{c}{2\zeta^2L_{\textrm{eff}}}\,.
\end{equation}
The effective cavity length
\begin{equation}
L_{\textrm{eff}}=2l+L(\sqrt{1+\zeta'^2}-\zeta')^2
\end{equation}
shows a decrease from $2l+L$ to $2l$, as the middle mirrors' polarizability $\zeta'$ is increased.

The well-resolved resonance condition can be found as previously by stipulating that $\kappa<c/2(L+2l)$. In the regime $\zeta\gg\zeta'\gg 1$ and $l\ll L$, this implies that
\begin{equation}
\frac{1}{4\zeta'^2}+\frac{2l}{L}>\frac{1}{\zeta^2}
\end{equation} has to be satisfied in order for a single-mode description to be valid. Outside this regime, tunneling occurs~\cite{Dombi2013} between two or more modes and the transmission spectrum does no longer allow for identifying single modes.

When $\zeta\gg 1$, one finds mean field intensities in the left/right and middle subcavities given by
\begin{equation}
|E_l|^2\simeq 2\zeta^2\,,
\end{equation} and
\begin{equation}
|E_L|^2\simeq 2\zeta^2(\sqrt{1+\zeta'^2}-\zeta')^2\,,
\end{equation}
respectively. We thus expect a similar behavior to that observed in the three-mirror case, i.e., a reduction in the field intensity in the middle cavity when $\zeta'$ increases, while the field intensity in the left/right subcavities remains at a level comparable to what it would be in a resonant symmetric cavity with mirror reflectivity $\zeta$.

\begin{figure}
\centering
\includegraphics[width=0.9\columnwidth]{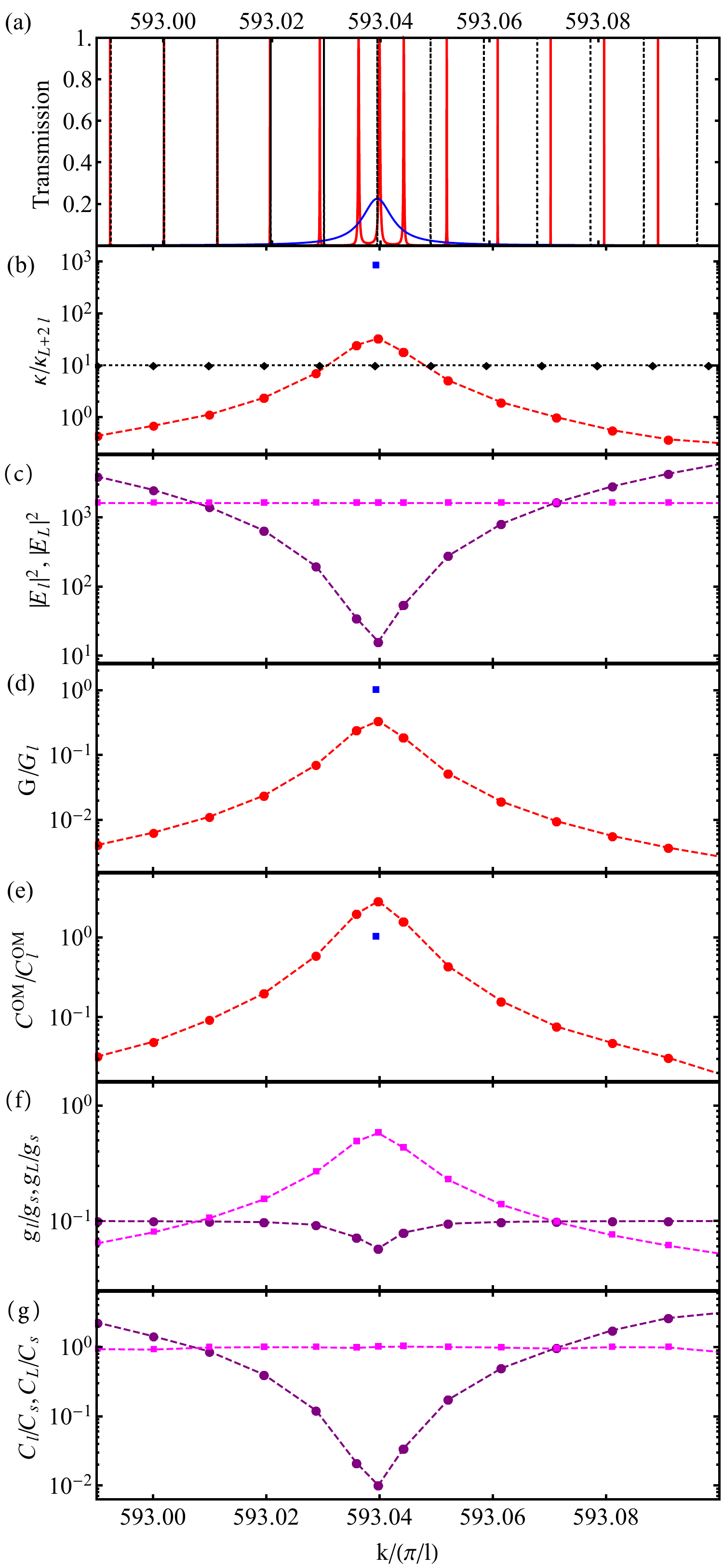}
\caption{(a) Transmission spectra (red) as a function of $k/(\pi/l)$ for a four-mirror cavity with $\zeta=20$, $\zeta'=5$, $L=100\pi$, $l=\pi$. The black and blue curves show the spectra of a cavity without middle mirrors and with length $L+2l$ and of the right subcavity with length $l$, respectively. (b) HWHM of the resonances shown in (a), normalized to that of a cavity without middle mirrors and with length $L+2l$. (c) Mean field intensities in middle (purple circles) and left/right subcavities (magenta squares), normalized to the incident field intensity. (d) Linear optomechanical coupling (red dots), normalized to that of the right subcavity with length $l$ (blue point). (e) Optomechanical cooperativity (red dots), normalized to that of the right subcavity with length $l$ (blue point). (f) Jaynes-Cummings coupling in the middle (purple circles) and left/right subcavities (magenta squares), normalized to the coupling $g_s$ of a cavity of length $l$ and mirror polarizability $\zeta$. (g) Corresponding normalized cooperativities.} \label{fig:fig6}
\end{figure}

This behavior is illustrated in Fig.~\ref{fig:fig6} in the case of a four-mirror cavity consisting in two short subcavities separated by a long middle cavity ($L/l=100$) and with middle mirrors having poorer reflectivity than the outer mirrors ($\zeta'=5$, $\zeta=20$). For resonances in the vicinity of the short subcavity resonance shown in Fig.~\ref{fig:fig6}a the linewidth is observed to increase (Fig.~\ref{fig:fig6}b), the field intensity remains high in the short cavities and drops in the long middle cavity (Fig.~\ref{fig:fig6}c).

In other words, by operating around such 'common' resonances for the compound system, it is possible to confine the field of a (long) symmetric cavity in the (short) left/right subcavities, with the field in the (long) middle cavity ultimately tending towards a value which corresponds to free-space propagation. This effect is in essence the opposite of that discussed in~\cite{Xuereb2012,Xuereb2013}, where resonances were chosen so as to \textit{enhance} the field intensity in between the intracavity reflectors, resulting in an effective narrowing of the linewidth of the compound resonator (see also~\cite{Nair2016}). We show in the next sections how to exploit this structural field confinement for optomechanical and cavity QED interactions.

\subsection{Optomechanics}\label{sec:OM}
To examine the strength of optomechanical interactions around such a common resonance, let us assume that one of the left/right subcavity end-mirror is movable and look at its dispersive optomechanical coupling with the cavity field. In a fashion similar to that used in Sec.~\ref{sec:three} one shows that
\begin{equation}
G=\frac{ck\zeta^2}{2l\zeta^2+L(\zeta\sqrt{1+\zeta'^2}-\zeta'\sqrt{1+\zeta^2})^2}\,,
\end{equation}
which, in the regime $\zeta\gg 1$, reduces to $G=ck/L_{\textrm{eff}}$, with the effective length$L_{\textrm{eff}}$ defined in~(\ref{eq:kappa4}). This analytical prediction is corroborated by the numerical results shown in Fig.~\ref{fig:fig6}d, where $G$ is seen to increase around the common resonance. As a consequence of the concomitant and equal increase of the linewidth and optomechanical coupling, the optomechanical cooperativity is also correspondingly increased and may become larger than what it would be in the right subcavity alone. Let us also mention that similar results are obtained if the movable mirror is one of the middle mirrors.

\subsection{Effective Jaynes-Cummings model}\label{sec:JC}
Let us now go back to a static mirror arrangement, but assume that a two-level system is positioned in one of the subcavities and couples to the field via a Jaynes-Cummings interaction. Under the assumptions listed in Sec.~\ref{sec:two}, we are interested in the effective mode volume defined via the energy relation (see e.g.~\cite{VanEnk1999} or~\cite{Lepert2011})
\begin{equation}
\int |E_{\textrm{vac}}|^2dV =
\frac{1}{2}\varepsilon_0A\left(\mathcal{E}_{0,L}^2L+2\mathcal{E}_{0,l}^2l\right)=\frac{1}{2}\hbar\omega\,.
\end{equation}
Since $g\propto \mathcal{E}_0$ and $\mathcal{E}_{0,L}/\mathcal{E}_{0,l}=E_L/E_l$ (up to a phase), one immediately gets that
\begin{subequations}
\begin{align}
g_l&=\frac{\beta}{\sqrt{2l+L|E_L/E_l|^2}}\,,\\
g_L&=\frac{\beta}{\sqrt{L+2l|E_l/E_L|^2}}\,,
\end{align}
\end{subequations}
with $\beta=(d/\hbar)\sqrt{\hbar\omega/\varepsilon_0A}$.
In the regime $\zeta\gg 1$ the maximal Jaynes-Cummings coupling strength $g_l$ for an emitter located in the left/right subcavities thus scales as $L_{\textrm{eff}}^{-1/2}$, with $L_{\textrm{eff}}$ the effective length introduced in Eq.~(\ref{eq:kappa4}). In the good cavity limit ($\zeta'\gg 1$), one has that $g_l\varpropto 1/\sqrt{2l}$, i.e., the quantum emitter couples to a delocalized field which is essentially confined within the two subcavities with length $l$. Let us note that, since the compound cavity linewidth is given by half of the linewidth of one subcavity, the Jaynes-Cummings cooperativity $C=g_l^2/\kappa \gamma$ remains at the same value as it would have in an isolated symmetric cavity with length $l$ and polarizability $\zeta$. This behavior can be observed in Figs.~\ref{fig:fig6}f~and~\ref{fig:fig6}g. Obviously, due to the confinement of the field in the left and right subcavities, the coupling strength and cooperativity in the middle cavity are concomitantly reduced.

\section{Modular hybrid optomechanics}\label{sec:modular}

In the previous section it was shown that operating in the vicinity of a common resonance of a four-mirror compound system allows for confining the field into the short subcavities, where light-matter interactions can be (i) enhanced if one compares with a situation of an equally long cavity void of middle mirrors or (ii) kept at similar strength levels if one compares with the situation of the short subcavities considered alone. This provides in principle a generic modular approach for modelling cavity networks containing, e.g., quantum emitters and/or optomechanical resonators, which is valid for any mirror reflectivity or spacing. It can be applied in particular to a modular hybrid atom-optomechanical interface in which a cavity containing quantum emitters is coupled to an optomechanical cavity via an optical link. A natural issue to consider is then the effect of the relative lengths of the subcavities on the interface performance in terms of coupling rates. Indeed, in a specific implementation, practical constraints may impose limitations on, e.g., the lengths that are required to achieve good optomechanical coupling with the movable element in the optomechanical cavity or to accommodate the quantum emitters in the cavity QED resonator.

\begin{figure}[H]
\centering
\includegraphics[width=0.92\columnwidth]{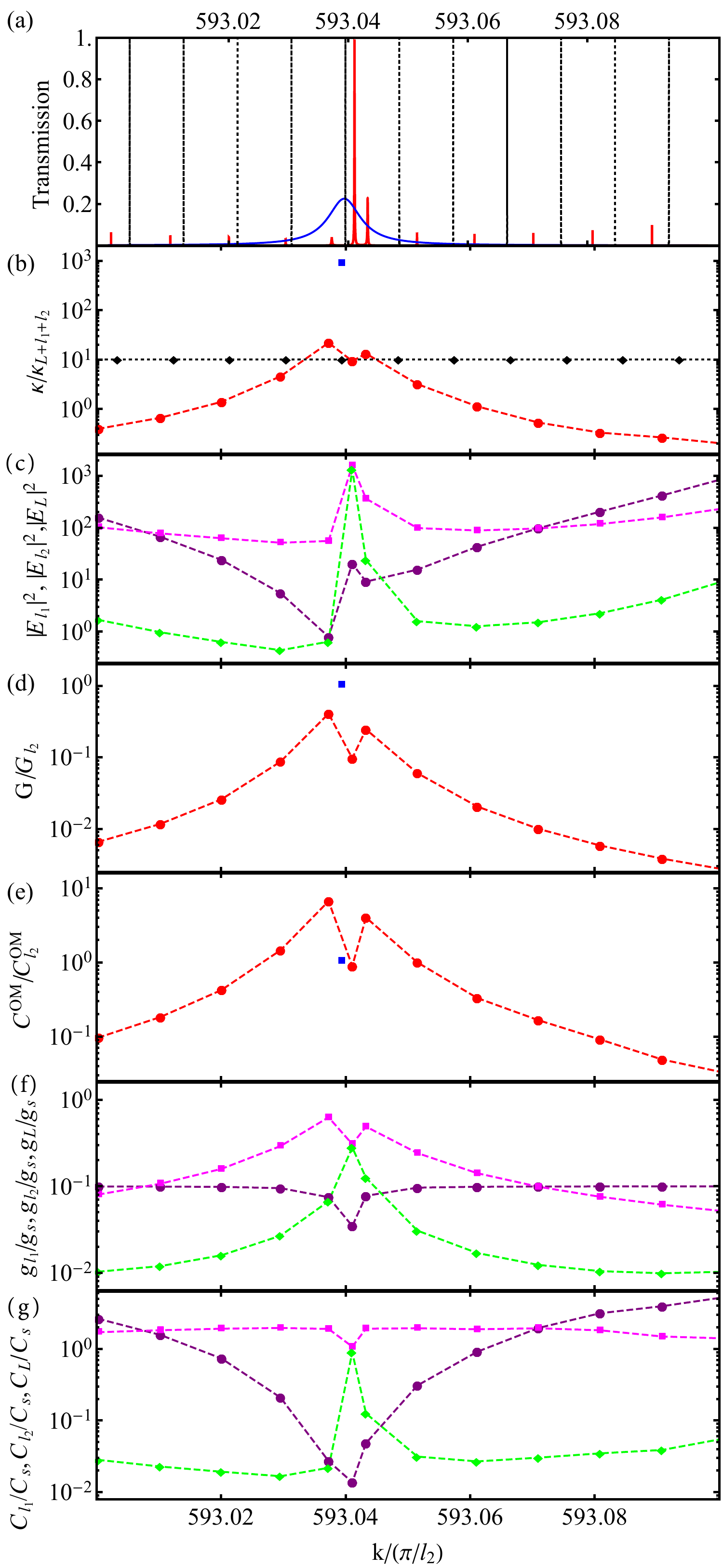}
\caption{(a) Transmission spectrum (red) as a function of $k/(\pi/l)$ for a four-mirror cavity with $\zeta=20$, $\zeta'=5$, $L=100\pi$, $l_1=9.91\pi$ and $l_2=\pi$. The black and blue curves show the spectra of a cavity without middle mirrors and with length $L+l_1+l_2$ and of the right subcavity with length $l_2$, respectively. (b) Normalized HWHM of the resonances shown in (a). (c) Mean field intensities in left (green losanges), middle (purple circles) and right (magenta squares) subcavities. (d) Linear optomechanical coupling (red dots), normalized to that of the right subcavity with length $l_2$ (blue point). (e) Optomechanical cooperativity (red dots), normalized to that of the right subcavity with length $l_2$ (blue point). (f) Jaynes-Cummings coupling in left (green losanges), middle (purple circles) and right (magenta squares) subcavities, normalized to the coupling $g_s$ of a cavity of length $l_2$ and mirror polarizability $\zeta$. (g) Corresponding normalized cooperativities.} \label{fig:fig7}
\end{figure}

We investigate this issue with a simple numerical example in a four-mirror configuration similar to the one studied previously, the mirror separations being now $l_1=9.91\pi$, $L=100\pi$, and $l_2=\pi$, respectively. The (larger) left subcavity with length $l_1$ accommodates a quantum emitter which couples to the field via a Jaynes-Cummings interaction, as discussed in Sec.~\ref{sec:JC}, while the (shorter) right subcavity with length $l_2$ has a movable end-mirror subjected to the radiation pressure of the field. The goal is thus to extract the parameters for which the dynamics of the compound system could be modelled by an interaction Hamiltonian
\begin{equation}
H=\hbar g_{l_1}(\hat{a}\hat{\sigma}_++\hat{a}^{\dagger}\hat{\sigma}_-)+\hbar G_{l_2}\hat{a}^{\dagger}\hat{a} \hat{x}\,
\end{equation}
where the field operators are associated with a single field mode spanning the whole compound cavity and decaying out of the cavity with the effective rate extracted from the transfer matrix calculations.

In this case, one can look for a $k$ vector such that resonance conditions (now non-degenerate) are simultaneously satisfied in all three cavities, and proceed as previously. When such a common resonance exists, one can derive similar analytical expressions for the effective length, linewidth and Jaynes-Cummings/optomechanical couplings, as in the previous section, if one substitutes $2l$ with $l_1+l_2$. This means, again, that one confines the field in the left and right subcavities, but with coupling strengths now scaling with the {\it total} length of both subcavities $l_1+l_2$. This is illustrated in Fig.~\ref{fig:fig7} if one looks at the resonance of the compound cavity closest to the optomechanical cavity resonance (unity transmission peak in Fig.~\ref{fig:fig7}a). As a consequence of the simultaneous quasi-resonance in all three cavities, a high overall transmission level is observed. Both the optomechanical and Jaynes-Cummings coupling strengths are reduced, though, as compared to Fig.~\ref{fig:fig6}, due to the increase in the value of $l_1+l_2$. The values of the cooperativities remain similar to those that would be obtained in the subcavities alone.

However, the previous working point need not be optimal. Indeed, due to multiple mirror interferences, it is possible to find resonances for which the field can be strongly confined in one subcavity, while being weaker in the other subcavities. This can be observed, for instance, for the resonance corresponding to the peak left to the unit transmission peak in Fig.~\ref{fig:fig7}a. Such a resonance shows a reduced overall transmission level, but a comparatively larger optomechanical cooperativity (about 8 times larger than it would be either in the optomechanical cavity alone or in the compound cavity at the hybridized resonance). As can be seen from the mean field intensities in the subcavities, a larger relative intensity difference is indeed obtained between the middle and the right subcavity, resulting in a larger radiation pressure force on the movable mirror in the right subcavity. The enhanced optomechanical interaction occurs, though, at the expense of a reduction in coupling with the quantum emitter in the left subcavity, since the field is comparatively weaker there. However, if this reduction can be somehow compensated in practice, e.g. by the collective coupling with an ensemble of emitters instead, then such a resonance might be preferable in terms of hybrid coupling performances.

\section{Conclusion}
We employed a one-dimensional transfer matrix approach to derive optical, optomechanical and quantum emitter-light interaction properties of idealized multi-element resonators. We provided a simple and general roadmap to derive common resonances defining spatially extended modes that can be used as a starting point for building quantum optics models for the dynamics of a variety of systems. 

In particular, we analyzed resonators systems of three and four elements widely present in (hybrid) optomechanical setups~\cite{Thompson2008,Wilson2009,Karuza2012,Kemiktarak2012NJP,Purdy2013,Flowers2012,Shkarin2014,Hammerer2009strong,Hammerer2009establishing,Hunger2010resonant,Camerer2011,Genes2011,Vogell2013,Vogell2015,Nair2016} or in systems of coupled (fiber-)cavities with atoms, NV-centers in diamond, quantum dots, molecules, etc.~\cite{Takahashi2013,Steiner2013,Brandstatter2013,Colombe2007,Becher2013,MiguelSanchez2013,Kelkar2015}. We provided simple analytical criteria for the emergence of tunneling and derived analytical expressions for the effective linewidth, optomechanical coupling strength and Jaynes-Cummings interaction strengths.

As a by-product of this analysis we show for example in the three-resonator system that enhanced optomechanical coupling interactions can be obtained in a membrane-at-the-end configuration.

Depending on the systems at hand and potential experimental constraints, such a roadmap thus provides a convenient and rigorous way to analyze and design multiple-element optical systems with non-trivial field distributions and optimize the performances of a given hybrid modular architecture.

\section*{Acknowlegments}
We acknowledge support from Aarhus Universitets Forskningsfond, the Danish National
Council for Independent Research, Villum Fonden and the Austrian Science Fund (FWF) via project P29318-N27.

\section{Appendix}

For completeness we give here the exact expressions for the HWHM of a common resonance defined by Eq.~(\ref{eq:res3}) in the three-mirror configuration:
\begin{widetext}
\begin{equation}
\kappa=\frac{c}{2}\sqrt{\frac{1+\zeta'^2}{\zeta\left((l^2+L^2)\zeta(1+\zeta^2)(1+2\zeta'^2)+2Ll\zeta(1+\zeta^2)+(l^2-L^2)\zeta'(1+2\zeta^2)\sqrt{(1+\zeta^2)(1+\zeta'^2)}\right)}}\,,
\end{equation}
and by Eq.~(\ref{eq:res4}) in the symmetric four-mirror configuration (left/right subcavities with equal length)
\begin{equation}
\kappa=\frac{c}{\sqrt{\left[L\left((1+2\zeta^2)(1+2\zeta'^2)-4\zeta\zeta'\sqrt{(1+\zeta^2)(1+\zeta'^2)}\right)-2l(1+2\zeta^2)\right]^2-(2l+L)^2-8lL\zeta'^2}}\,.
\end{equation}
\end{widetext}

\bibliography{mod_bib}

\end{document}